\documentclass[onecolumn, 12pt]{IEEEtran}

\ifCLASSINFOpdf
\else
\fi

\usepackage{bm}
\usepackage{color}
\usepackage{float}
\usepackage{amsthm}
\usepackage{amsmath}
\usepackage{amssymb}
\usepackage{amsfonts}
\usepackage{fancyhdr}
\usepackage{graphicx}
\usepackage{placeins}
\usepackage{enumerate}
\usepackage{extarrows}
\usepackage{subfigure}
\usepackage{algorithmic}
\usepackage{tabularx,colortbl}
\usepackage[numbers,sort&compress]{natbib}
\usepackage{setspace}

\newtheoremstyle{mytheorem}
{0pt} 
{0pt} 
{\normalfont} 
{1em} 
{\em} 
{:} 
{0.5em} 
{} 
\theoremstyle{mytheorem}

\newtheorem{mytheorem}{Theorem}
\newtheorem{myremark}{Remark}

\begin{document}

\title{Spectral and Energy Efficiency of IRS-Assisted MISO Communication with Hardware Impairments}

\author{
Shaoqing~Zhou,
Wei~Xu,~\IEEEmembership{Senior Member,~IEEE,}
Kezhi~Wang,~\IEEEmembership{Member,~IEEE,} \\
Marco~Di~Renzo,~\IEEEmembership{Fellow,~IEEE,}
and~Mohamed-Slim~Alouini,~\IEEEmembership{Fellow,~IEEE}

\thanks{S. Zhou is with the National Mobile Communications Research Laboratory, Southeast University, Nanjing 210096, China (e-mail: sq.zhou@seu.edu.cn).}
\thanks{W. Xu is with the National Mobile Communications Research Laboratory, Southeast University, Nanjing 210096, China, and also with Purple Mountain Laboratories, Nanjing 211111, China (e-mail: wxu@seu.edu.cn).}
\thanks{K. Wang is with the Department of Computer and Information Sciences, Northumbria University, Newcastle upon Tyne NE1 8ST, U.K. (e-mail: kezhi.wang@northumbria.ac.uk).}
\thanks{M. Di Renzo is with Universit\'e Paris-Saclay, CNRS and CentraleSup\'elec, Laboratoire des Signaux et Syst\`emes, Gif-sur-Yvette, France. (e-mail: marco.direnzo@centralesupelec.fr).}
\thanks{M.-S. Alouini is with the Division of Computer, Electrical, and Mathematical Science and Engineering, King Abdullah University of Science and Technology, Thuwal 23955-6900, Saudi Arabia (e-mail: slim.alouini@kaust.edu.sa).}
}

\maketitle

\vspace{-5mm}

\begin{abstract}
In this letter, we analyze the spectral and energy efficiency of an intelligent reflecting surface (IRS)-assisted multiple-input single-output (MISO) downlink system with hardware impairments. An extended error vector magnitude (EEVM) model is utilized to characterize the impact of radio-frequency (RF) impairments at the access point (AP) and phase noise is considered for the imperfect IRS. We show that the spectral efficiency is limited due to the hardware impairments even when the numbers of AP antennas and IRS elements grow infinitely large, which is in contrast with the conventional case with ideal hardware. Moreover, the performance degradation at high SNR is shown to be mainly affected by the AP hardware impairments rather than the phase noise of IRS. We further obtain the optimal transmit power in closed form for energy efficiency maximization. Simulation results are provided to verify these results.
\end{abstract}

\vspace{-5mm}

\begin{IEEEkeywords}
Intelligent reflecting surface, hardware impairments, downlink spectral efficiency, energy efficiency.
\end{IEEEkeywords}

\IEEEpeerreviewmaketitle

\vspace{-5mm}

\section{Introduction}
\IEEEPARstart{I}{ntelligent} reflecting surface (IRS) has recently been acknowledged as a promising new technology to realize spectral-, energy- and cost-efficient wireless communication for the fifth generation network and beyond \cite{Dang2020What}. IRS is a planar array consisting of a large number of low-cost reflecting elements, which independently induce phase adjustments on impinging signals to conduct reflecting beamforming. Significantly different from existing technologies, IRS reconfigures wireless communication environment between transmitter and receiver via programmable and highly controllable intelligent reflection. Moreover, it avoids active radio-frequency (RF) chains and operates passively for short range coverage enhancement so that it can be densely deployed in a flexible way with affordable hardware cost and energy consumption.

Traditional communication theories may no longer be applied because the IRS-assisted wireless system consists of both active and passive components, instead of solely active entities \cite{Renzo2019Smart}. Researches on channel estimation, IRS beamforming design and system performance analysis are on the way. Two efficient uplink channel estimation schemes were proposed in \cite{Zheng2020Intelligent} for IRS-assisted multi-user systems with various channel setups. In \cite{Pan2019Intelligent}, transmit precoding and passive IRS phase shifts were jointly optimized for simultaneous wireless information and power transfer systems. Ergodic spectral efficiency of an IRS-assisted massive multiple-input multiple-output system was characterized in \cite{Han2019Large} under Rician fading channel. In \cite{Gao2020Reconfigurable}, spectral efficiency of an IRS-aided multi-user system was studied under proportional rate constraints and an iteratively optimizing solution was proposed with closed-form expressions. Secrecy rate was maximized in \cite{Shen2019Secrecy} for an IRS-assisted multi-antenna system by alternately optimizing transmitting covariance and IRS phase shifts. IRS was also shown to be effective in enhancing the performance of cell-edge users \cite{Pan2019Multicell}.

In practice, precise phase control is infeasible at IRS due to hardware limitations and imperfect channel estimation. Corresponding researches are still in their infancy. Discrete phase shifts were considered for IRS-assisted multi-user communication in \cite{Di2019Hybrid}, where a hybrid beamforming optimization algorithm was proposed for sum rate maximization. In addition to non-ideal IRS, the impacts of RF impairments at transmitter on the performance of an IRS system have not been clear. To capture the aggregate impacts of various RF impairments, a generalized model named extended error vector magnitude (EEVM) was proposed in \cite{Schenk2008RF} for cellular transmitters.

In this letter, we focus on an IRS-assisted multiple-input single-output (MISO) system with hardware impairments at both access point (AP) and IRS. Theoretical expression of spectral efficiency is derived for this non-ideal case. We discover that the performance is limited even with increasing numbers of elements at both the AP and IRS. The impact of phase noise at IRS diminishes at high SNR. Meanwhile, we obtain a closed-form solution to the optimal power design for maximizing energy efficiency. The optimal power increases with RF impairments.

\vspace{-5mm}

\section{System Model}

\vspace{-3mm}

\subsection{Signal Model}
We consider a MISO downlink system where an IRS consisting of $N$ reflecting elements is deployed to assist the communication from an $M$-antenna AP to a single-antenna user. The IRS is triggered by an attached smart controller connected to the AP. Denote the reflection matrix of IRS by $\bm{\Theta} = \text{diag}\{\zeta_1 e^{j\theta_1}, \zeta_2 e^{j\theta_2}, \dots, \zeta_N e^{j\theta_N} \}$, where $\zeta_n \in [0, 1]$ and $\theta_n \in [0, 2\pi)$ for $n = 1, 2, \dots, N$ are respectively the amplitude reflection coefficient and the phase shift introduced by the $n$th reflecting element. In practice, each reflecting element is usually designed to maximize the signal reflection. Without loss of generality, we set $\zeta_n = 1$ for all $n$ \cite{Zheng2019Intelligent}. The direct link between AP and user is blocked by obstacles, such as buildings or human body, which is common in the communication at high-frequency bands, like millimeter wave. Thus it would be better to deploy IRS at positions where line-of-sight (LoS) communication is ensured for both AP-to-IRS and IRS-to-user links.

Considering the flat-fading model, the channel from the AP to IRS and that from the IRS to user are respectively denoted by $\mathbf{H}_1$ and $\mathbf{h}_2^H$. Both channels are assumed to be LoS, which are represented by
\begin{equation}
\mathbf{H}_1 = \alpha \mathbf{a}_N(\phi_r^a, \phi_r^e) \mathbf{a}^H_M(\phi_t^a, \phi_t^e),\ \mathbf{h}_2^H = \beta \mathbf{a}^H_N(\varphi_t^a, \varphi_t^e),
\end{equation}
where $\alpha$ and $\beta$ are the corresponding strength of path AP-to-IRS and IRS-to-user, $\phi_r^a$ ($\phi_r^e$) is the azimuth (elevation) angle of arrival (AoA) at IRS, $\phi_t^a$ ($\phi_t^e$) and $\varphi_t^a$ ($\varphi_t^e$) are the azimuth (elevation) angles of departure (AoD) at AP and IRS, respectively, and $\mathbf{a}_{X}(\vartheta^a, \vartheta^e)$ is the array response vector. We consider uniform square planar array (USPA) with $\sqrt{X} \times \sqrt{X}$ antennas. The array response vector can be written~as
\begin{equation}
\mathbf{a}_X(\vartheta^a, \vartheta^e) = [ 1, \dots, e^{j2\pi\frac{d}{\lambda}(x \sin\vartheta^a \sin\vartheta^e + y \cos\vartheta^e)}, \dots, e^{j2\pi\frac{d}{\lambda}((\sqrt{X}-1) \sin\vartheta^a \sin\vartheta^e + (\sqrt{X}-1) \cos\vartheta^e)} ]^T,
\end{equation}
where $d$ and $\lambda$ are the antenna spacing and signal wavelength, and $0 \leq x, y < \sqrt{X}$ are the antenna indices in the planar. Assume that the AP knows the channel state information (CSI) of both $\mathbf{H}_1$ and $\mathbf{h}_2^H$. Channel estimation methods for communication with IRS can be found in \cite{Zheng2019Intelligent}\cite{Ning2019Channel}.

With the errors caused by imperfect RF chains at AP, we adopt the EEVM in \cite{Schenk2008RF} to model the transmit signal, which can be written as
\begin{equation}
\mathbf{x} = \bm{\chi} \mathbf{w} s + \mathbf{n}_{RF},
\end{equation}
where $s$ is the signal satisfying $\mathbb{E} [ |s|^2 ] = P$ with $P$ being the transmit power budget, $\mathbf{w}$ is the normalized beamforming vector at AP, $\bm{\chi} = \text{diag}\{\chi(1), \chi(2), \dots, \chi(M)\}$ with $\chi(m) = \eta(m) e^{j\psi(m)}$ for $m = 1, 2, \dots, M$ representing the RF attenuation and phase rotation of the $m$th RF chain with $|\eta(m)| \leq 1$, and $\mathbf{n}_{RF} = [ n_{RF}(1), n_{RF}(2), \dots, n_{RF}(M) ]^T$ represents the additive distortion noise with covariance matrix $\mathbf{C}_{\mathbf{n}_{RF}}$. The mapping of $\bm{\chi}$ and $\mathbf{n}_{RF}$ to particular type(s) of RF impairments, e.g., phase noise, I/Q imbalance and nonlinearity, could be found in\setcitestyle{open=[,close=}\cite{Schenk2008RF}, Ch. 7]. For notational simplicity, assume that $\psi(m)$ is uniformly distributed as $\mathcal{U}[-\delta_{\psi(m)}, \delta_{\psi(m)}]$ with $\delta_{\psi(m)} \in [0, \pi)$, $n_{RF}(m) \sim \mathcal{CN} (0, \sigma^2(m))$, and the impairments of all RF chains fall into the same level, i.e., $\eta(m) = \eta$, $\delta_{\psi(m)} = \delta_{\psi}$, $\sigma(m) = \sigma$ and $\mathbf{C}_{\mathbf{n}_{RF}} = \sigma^2 \mathbf{I}_M$.

Furthermore, there always exist some phase errors at IRS in implementation. The received signal with phase errors can be modeled as
\begin{equation} \label{y}
y = \mathbf{h}_2^H \widetilde{\bm{\Theta}} \mathbf{H}_1 \mathbf{x} + u = \mathbf{h}_2^H \widetilde{\bm{\Theta}} \mathbf{H}_1 \bm{\chi} \mathbf{w} s + \mathbf{h}_2^H \widetilde{\bm{\Theta}} \mathbf{H}_1 \mathbf{n}_{RF} + u,
\end{equation}
where $\widetilde{\bm{\Theta}} = \text{diag}\{ e^{j\tilde{\theta}_1}, e^{j\tilde{\theta}_2}, \dots, e^{j\tilde{\theta}_N} \}$ with $\tilde{\theta}_n = \theta_n + \hat{\theta}_n$ being the practical phase shift of the $n$th reflecting element, $\hat{\theta}_n$ is the phase noise due to the fact, e.g., only discrete phase shifts are possible at IRS, and $u$ is the additive noise with zero mean and variance $\sigma^2_u$. Assume that $\hat\theta_n$ is uniformly distributed as $\mathcal{U}[-\delta_{\hat{\theta}}, \delta_{\hat{\theta}}]$ with $\delta_{\hat{\theta}} \in [0, \pi)$. Since the distortion noise is independent of channel noise, the received SNR is given by
\begin{equation} \label{SNR}
\text{SNR} = \frac{P | \mathbf{h}_2^H \widetilde{\bm{\Theta}} \mathbf{H}_1 \bm{\chi} \mathbf{w} |^2}{(\mathbf{h}_2^H \widetilde{\bm{\Theta}} \mathbf{H}_1) \mathbf{C}_{\mathbf{n}_{RF}} (\mathbf{h}_2^H \widetilde{\bm{\Theta}} \mathbf{H}_1)^H + \sigma^2_u}.
\end{equation}
Then we have the downlink spectral efficiency as
\begin{equation} \label{R}
R = \log_2 (1 + \text{SNR}).
\end{equation}

\vspace{-8mm}

\subsection{Power Consumption Model}
Before we discuss the power consumption, it needs to be emphasized that the IRS does not consume any transmit power due to its nature of passive reflection. The total power consumption is modeled as~\setcitestyle{open=[,close=]}\cite{Huang2019Reconfigurable}
\begin{equation} \label{Ptotal}
P_{\text{T}} = \mu P + P_{\text{C}},
\end{equation}
where $\mu = \nu^{-1}$ with $\nu$ being the efficiency of transmit power amplifier considering the RF impairments and $P_{\text{C}}$ is the total static hardware power dissipated in all circuit blocks. The establishment of (\ref{Ptotal}) models well under two assumptions: 1) the transmit amplifier operates within its linear region; and 2) the power consumption $P_{\text{C}}$ does not rely on the rate of the communication link. Both assumptions are valid in typical wireless systems.

\vspace{-5mm}

\section{Spectral and Energy Efficiency Analysis}
In this section, we quantitatively analyze the downlink spectral and energy efficiency and discover the impact of hardware impairments at both the AP and IRS. The ideal spectral and energy efficiency are retrieved as a special case of our analysis and it is presented for comparison.

\vspace{-5mm}

\subsection{Spectral Efficiency Analysis}
Before analyzing the performance, we need to determine the transmit beamforming of AP and the reflecting beamforming of IRS. Since the hardware impairments are unknown and in order to facilitate the design in practice, the two parameters, $\mathbf{w}$ and $\bm{\Theta}$, are optimized by treating the hardware as ideal. Maximum ratio transmission (MRT) is adopted for transmit beamforming, i.e.,
\begin{equation} \label{w}
\mathbf{w} = (\mathbf{h}_2^H \bm{\Theta} \mathbf{H}_1)^H / \Vert \mathbf{h}_2^H \bm{\Theta} \mathbf{H}_1 \Vert.
\end{equation}
We identify the optimal reflecting beamforming of IRS by maximizing the received signal power as
\begin{align} \label{Thetaopt} \nonumber
\bm{\Theta}^{opt} &= \arg \max_{\bm{\Theta}} |\mathbf{h}_2^H \bm{\Theta} \mathbf{H}_1 \mathbf{w}|^2 \overset{(a)}= \arg \max_{\bm{\Theta}} \Vert \mathbf{h}_2^H \bm{\Theta} \mathbf{H}_1 \Vert^2 \\ \nonumber
&= \arg \max_{\bm{\Theta}} | \mathbf{a}^H_N(\varphi_t^a, \varphi_t^e) \bm{\Theta} \mathbf{a}_N(\phi_r^a, \phi_r^e) |^2 \Vert \mathbf{a}^H_M(\phi_t^a, \phi_t^e) \Vert^2 \\
&\overset{(b)}= \arg \max_{\bm{\Theta}} |\sum_{\substack{0 \leq x, y < \sqrt{N}, \\ n = \sqrt{N}x+y+1}} e^{j2\pi\frac{d}{\lambda}(xp + yq) + j \theta_n} |^2,
\end{align}
where $(a)$ is obtained by substituting $\mathbf{w}$ in (\ref{w}), $(b)$ makes use of a mapping from the two-dimensional index $(x, y)$ to the index $n$, $p = \sin\phi_r^a \sin\phi_r^e - \sin\varphi_t^a \sin\varphi_t^e$, and $q = \cos\phi_r^e - \cos\varphi_t^e$. Observing (\ref{Thetaopt}), it is easy to get the optimal phase shift of the $n$th reflecting element as
\begin{equation} \label{thetaopt}
\theta_n^{opt} = - 2 \pi \frac{d}{\lambda} (xp + yq),
\end{equation}
where $x = \lfloor (n-1)/\sqrt{N} \rfloor$ and $y = {(n-1)\ \text{mod}\ \sqrt{N}}$, and $\lfloor \cdot \rfloor$ represents rounding down the value and mod means taking the remainder.

Now considering the design of $\bm{\Theta}^{opt}$ in (\ref{thetaopt}) and $\mathbf{w}(\bm{\Theta}^{opt})$ in (\ref{w}), we characterize the impacts of both RF impairments at AP and phase noise at IRS on the downlink spectral efficiency in the following \emph{Theorem~1}.

\begin{mytheorem} \label{mytheorem1}
The downlink spectral efficiency for the massive IRS-assisted MISO with large $M$ and $N$ approaches
\begin{equation} \label{Rresult}
R \xlongrightarrow{a.s.} \log_2 \left( 1 + \frac{ P M N^2 \eta^2 |\alpha \beta|^2 \text{sinc}^2(\delta_{\psi}) \text{sinc}^2(\delta_{\hat{\theta}})}{M N^2 |\alpha \beta|^2 \text{sinc}^2(\delta_{\hat{\theta}}) \sigma^2 + \sigma^2_u} \right).
\end{equation}
\end{mytheorem}

\textbf{Special Case 1:} For the ideal system without any impairments, we let $\eta = 1$, $\sigma = 0$ and $\delta_{\psi} = \delta_{\hat{\theta}} = 0$ in (\ref{Rresult}). The downlink spectral efficiency reduces to
\begin{equation} \label{Rideal}
R_{ideal} = \log_2 \left(1 + \frac{P}{\sigma^2_u} M N^2 |\alpha \beta|^2 \right).
\end{equation}

\textbf{Special Case 2:} For high SNR, (\ref{Rresult}) in \emph{Theorem 1} can be further simplified as
\begin{equation} \label{Rresult2}
R \rightarrow \log_2 P + 2\log_2 \eta + 2\log_2 \text{sinc}(\delta_{\psi}) - \log_2 \sigma^2.
\end{equation}

\begin{myremark} \label{remark1}
It is concluded from (\ref{Rresult}) that the non-ideal spectral efficiency increases with $\eta$ while decreases with parameters $\delta_{\psi}$, $\sigma^2$ and $\delta_{\hat{\theta}}$. The impact of the phase rotation at AP in terms of $\delta_{\psi}$ is in general more significant than that of the phase noise at IRS in terms of $\delta_{\hat{\theta}}$.
\end{myremark}

\begin{myremark} \label{remark2}
The spectral efficiency in (\ref{Rresult}) increases with the transmit power approximately in a logarithmic manner similar to the ideal case in (\ref{Rideal}) but with a different scale. Contrary to the ideal case, the performance is ultimately upper bounded for increasing $M$ and $N$, which is $R(M, N) \leq \bar{R} = \log_2 (1 + \eta^2 \frac{P}{\sigma^2} \text{sinc}^2(\delta_{\psi}))$ for all large $M$ and $N$.
\end{myremark}

\begin{myremark} \label{remark3}
An interesting observation from (\ref{Rresult2}) is that the spectral efficiency at high SNR is merely limited by the RF impairments at AP rather than the phase noise at IRS, which can be explained from the perspective that the IRS reflecting beamforming simultaneously affects both the desired signal and the distortion noise under the considerations of hardware impairments at AP and LoS channel. It encourages us to use cheap IRS with low-resolution phase shifts without much consideration of performance degradation for large IRS.
\end{myremark}

\vspace{-5mm}

\subsection{Energy Efficiency Analysis}
The energy efficiency is defined as the ratio of the spectral efficiency to the power consumption, i.e., $\text{EE} \triangleq B R / P_{\text{T}}$ where $B$ is the channel bandwidth. We are interested in the performance at high SNR, which can be rewritten as
\begin{equation} \label{EE}
\text{EE} = \frac{B ( \log_2 P + 2\log_2 \eta + 2\log_2 \text{sinc}(\delta_{\psi}) - \log_2 \sigma^2 )}{\mu P + P_{\text{C}}}.
\end{equation}
In the following \emph{Theorem 2}, we give a closed-form expression of the optimal transmit power maximizing the EE in (\ref{EE}).

\begin{mytheorem} \label{mytheorem2}
The optimal transmit power to maximize the energy efficiency is the unique solution as
\begin{equation} \label{Popt}
P^{\star} = \mu P_{\text{C}}^{-1} W(\mu^{-1} e^{C_{\text{AP}}-1} P_{\text{C}}),
\end{equation}
where $W(x)$ is the Lambert's W-function and $C_{\text{AP}} = 2\ln \eta + 2\ln \text{sinc}(\delta_{\psi}) - \ln \sigma^2$.
\end{mytheorem}

Note that for an ideal system without hardware impairments, the optimal transmit power can be similarly derived as
\begin{equation}
P^{\star}_{ideal} = \mu P_{\text{C}}^{-1} W(\mu^{-1} e^{C - 1} P_{\text{C}}),
\end{equation}
where $C = \ln(M N^2 |\alpha \beta|^2) - \ln \sigma^2_u$. We emphasize that the IRS has continuous phase in the ideal case, which increases the static hardware power consumption of IRS.

\begin{myremark} \label{remark4}
For $P^{\star}$ in (\ref{Popt}) and $\text{EE}(P^{\star})$ in (\ref{EE}), the optimal transmit power increases with more severe RF impairments and the corresponding optimal energy efficiency decreases.
\end{myremark}

\vspace{-5mm}

\section{Simulation Results}
In this section, simulation results are presented to validate the results in Section \uppercase\expandafter{\romannumeral3}. Assume that $M = 16$, $N = 64$, $\eta = 0.9$, $\delta_{\psi} = \frac{\pi}{18}$, $\sigma^2 = 0.1$, $\delta_{\hat{\theta}} = \frac{\pi}{8}$, $\alpha = 0.1$, $\beta = 0.5$ and $\mu = 1.1$.

\begin{figure}
  \vspace{-8mm}
  \centering
  \includegraphics[width = 0.66\textwidth]{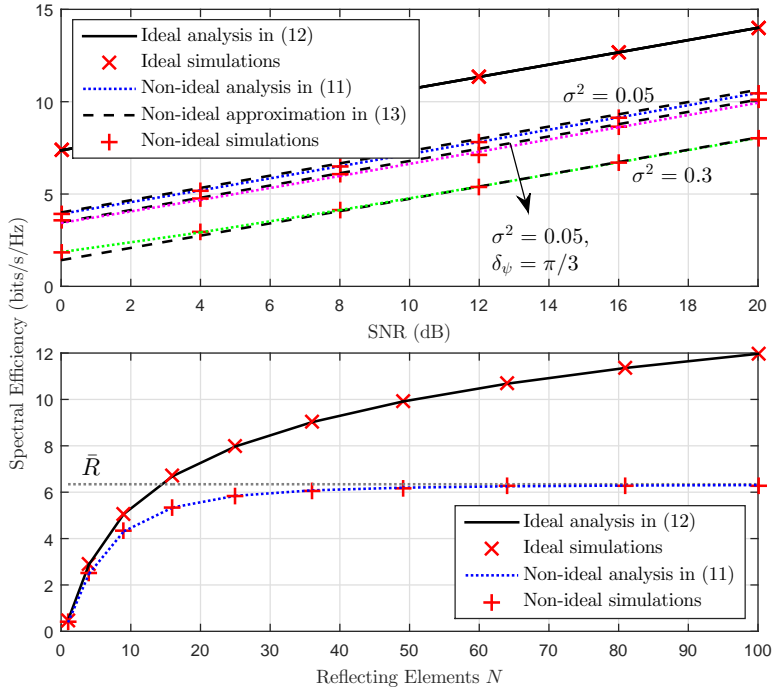}
  \vspace{-3mm}
  \caption{Downlink spectral efficiency versus SNR and $N$.}
  \label{Fig:simulationfig1}
  \vspace{-3mm}
\end{figure}

We plot the downlink spectral efficiency in \emph{Theorem 1}, special cases and by simulations in Fig. \ref{Fig:simulationfig1}. Both non-ideal case in (\ref{Rresult}) and ideal case in (\ref{Rideal}) increase with the transmit power but by respective scales. The simplified expression in (\ref{Rresult2}) appears to be fairly tight at high SNR.

We further assume SNR = $\frac{P}{\sigma^2}$ = 10 dB. Fig. \ref{Fig:simulationfig1} shows the spectral efficiency versus the number of IRS reflecting elements. As the number goes larger, the hardware impairments lead to limited growth of spectral efficiency, which is consistent with \emph{Remark 2}, while the ideal case continues increasing logarithmically with the squared number of elements.

\begin{figure}
  \vspace{-8mm}
  \centering
  \includegraphics[width = 0.5\textwidth]{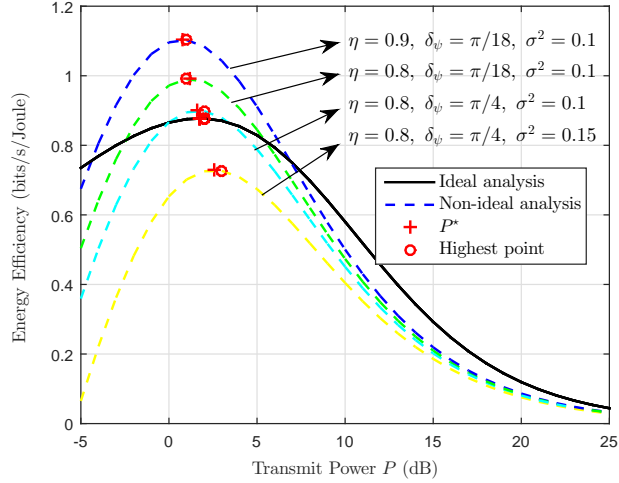}
  \caption{Energy efficiency versus $P$.}
  \label{Fig:simulationfig4}
\end{figure}

In Fig. \ref{Fig:simulationfig4}, we give the energy efficiency with various degrees of RF impairments. The optimal transmit power derived in \emph{Theorem 2} matches the highest point of the curve well. When the RF impairments become worse, higher optimal transmit power is required while the corresponding energy efficiency decreases. Note that the ideal case may obtain poorer performance than the non-ideal case because of larger static hardware power consumption of continuous-phase IRS.

\vspace{-5mm}

\section{Conclusion}
In this letter, we demonstrate the downlink spectral and energy efficiency of an IRS-assisted system with hardware impairments. The non-ideal spectral efficiency is upper bounded for large numbers of AP antennas and IRS elements. Specially, the impact of imperfect IRS diminishes at high SNR. The optimal transmit power for maximizing the energy efficiency increases as the RF impairments become more severe.

\appendices

\vspace{-5mm}

\section{Proof of Theorem 1}
Applying (\ref{w}) and (\ref{thetaopt}), we can rewrite the downlink spectral efficiency in (\ref{R}) as
\begin{equation} \nonumber
R = \log_2 \left( 1 + \frac{P | \mathbf{h}_2^H \widetilde{\bm{\Theta}} \mathbf{H}_1 \bm{\chi} (\mathbf{h}_2^H \bm{\Theta} \mathbf{H}_1)^H|^2 / \Vert \mathbf{h}_2^H \bm{\Theta} \mathbf{H}_1 \Vert^2}{\Vert \mathbf{h}_2^H \widetilde{\bm{\Theta}} \mathbf{H}_1 \Vert^2 \sigma^2 + \sigma^2_u} \right)
\end{equation}
\begin{equation}
\begin{split}
R &\overset{(c)}=  \log_2 \left( 1 + \frac{P |\alpha \beta|^2 \left|\sum\limits^N_{n = 1} e^{j\hat{\theta}_n}\right|^2 |\mathbf{a}^H_M(\phi_t^a, \phi_t^e) \bm{\chi} \mathbf{a}_M(\phi_t^a, \phi_t^e)|^2}{M \left(|\alpha \beta|^2 \left|\sum\limits^N_{n = 1} e^{j\hat{\theta}_n}\right|^2 \Vert \mathbf{a}_M(\phi_t) \Vert^2 \sigma^2 + \sigma^2_u \right) } \right) \\ \label{Rcal}
&= \log_2 \left( 1 + \frac{ P \eta^2 |\alpha \beta|^2 \left|\sum^M_{m = 1} e^{j\psi(m)}\right|^2 \left|\sum^N_{n = 1} e^{j\hat{\theta}_n}\right|^2 / M}{M |\alpha \beta|^2 \left|\sum^N_{n = 1} e^{j\hat{\theta}_n}\right|^2 \sigma^2 + \sigma^2_u} \right),
\end{split}
\end{equation}
where $(c)$ is obtained by substituting the equations $\mathbf{h}_2^H \bm{\Theta} \mathbf{H}_1 = \alpha \beta N \mathbf{a}^H_M(\phi_t^a, \phi_t^e)$ and $\mathbf{h}_2^H \widetilde{\bm{\Theta}} \mathbf{H}_1 = \alpha \beta \times$ $\sum^N_{n = 1} e^{j\hat{\theta}_n} \mathbf{a}^H_M(\phi_t^a, \phi_t^e)$.

For large $M \rightarrow \infty$, we have
\begin{equation} \label{mpsi}
\left| \frac{1}{M} \sum^M_{m = 1} e^{j\psi(m)} \right|^2 \xlongrightarrow[a.s.]{(d)} |\mathbb{E}[e^{j\psi(m)}]|^2 \overset{(e)}= |\mathbb{E}[\cos\psi(m)]|^2 \overset{(f)}= \text{sinc}^2(\delta_{\psi}),
\end{equation}
where $(d)$ applies the Strong Law of Large Numbers and the Continuous Mapping Theorem\setcitestyle{open=[,close=]}\cite{Xu2020Secure} which indicates that the convergence preserves for continuous matrix functions, $(e)$ uses the symmetry of the odd function $\sin\psi(m)$ for $\psi(m) \in [-\delta_{\psi}, \delta_{\psi}]$, $(f)$ is obtained by substituting the probability density function of variable $\psi(m)$, i.e., $f_X(x) = \frac{1}{2\delta_{\psi}}$ for $x \in [-\delta_{\psi}, \delta_{\psi}]$, and $\text{sinc}(x) = \frac{\sin x}{x}$. Similarly, for large $N \rightarrow \infty$, we have
\begin{equation} \label{ntheta}
\left|\frac{1}{N} \sum^N_{n = 1} e^{j\hat{\theta}_n}\right|^2 \xlongrightarrow{a.s.} \text{sinc}^2(\delta_{\hat{\theta}}).
\end{equation}
Substituting (\ref{mpsi}) and (\ref{ntheta}) into (\ref{Rcal}) completes the proof.

\vspace{-5mm}

\section{Proof of Theorem 2}
By calculating the partial derivative of EE in (\ref{EE}), we have
\begin{equation}
\frac{\partial}{\partial P}\text{EE} = B \frac{P^{-1} (\mu P + P_{\text{C}}) - \mu (\ln P + C_{\text{AP}})}{(\ln2)(\mu P + P_{\text{C}})^2}.
\end{equation}
Letting the partial derivative be zero, we have
\begin{align} \label{equ}
& \mu P (\ln P + C_{\text{AP}} -1 ) = P_{\text{C}}, \\ \nonumber
\xLeftrightarrow{t = \ln P}\ & \mu e^{t+C_{\text{AP}}-1} (t + C_{\text{AP}} -1 ) = e^{C_{\text{AP}}-1} P_{\text{C}}, \\ \label{t}
\overset{(g)}\Rightarrow \quad \ & t = W(\mu^{-1} e^{C_{\text{AP}} - 1} P_{\text{C}}) - C_{\text{AP}} + 1,
\end{align}
where $(g)$ uses the fact the Lambert's W-function is the inverse function of $f(W) = W e^W$. Now rearranging (\ref{t}) yields (\ref{Popt}).

The remainder proves that (\ref{equ}) has a unique solution. Define $g(P) \triangleq \mu P (\ln P + C_{\text{AP}} -1 )$. It follows $\frac{\mathrm{d}}{\mathrm{d}P}g(P) = \mu (\ln P + C_{\text{AP}}) > 0$.
Thus $g(P)$ is monotonically increasing with respect to $P$, which implies that equation (\ref{equ}) has at most one solution, which is exactly (\ref{Popt}).

\end{document}